\documentclass{elsart}
\usepackage{epsfig}
\usepackage{amssymb}
\usepackage{citesort}

\journal{Physics Letters B}
\def\v#1{\mbox{\boldmath$#1$}}
\def\ket#1{|#1 \rangle}
\def\bra#1{\langle #1|}

\begin{document}

\begin{frontmatter}

\title{Exchange terms in the two--nucleon induced non--mesonic
weak decay of $\Lambda$--hypernuclei}

\author{E. Bauer$^{1,2}$ and G. Garbarino$^3$}

\address{$^1$Departamento de F\'{\i}sica, Universidad Nacional de La Plata,\\
C. C. 67, 1900 La Plata, Argentina}

\address{$^2$Instituto de F\'{\i}sica La Plata,
CONICET, 1900 La Plata, Argentina}

\address{$^3$Dipartimento di Fisica Teorica, Universit\`a di Torino,\\
I-10125 Torino, Italy}

\date{\today}

\begin{abstract}
The contribution of Pauli exchange terms to the
two--nucleon induced non--mesonic weak decay of $^{12}_\Lambda$C
hypernuclei, $\Lambda NN\to nNN$ ($N=n$ or $p$),
is studied within a nuclear matter formalism implemented
in a local density approximation. We have adopted
a weak transition potential including the exchange of the complete
octets of pseudoscalar and vector mesons as well as a
residual strong interaction modeled on the Bonn potential. The introduction of
exchange terms turns out to reduce the two--nucleon induced non--mesonic
rate by 18\% and, jointly with an increase in the one--nucleon induced
rate by the same magnitude, reveals to be significant for an accurate
determination of the full set of hypernuclear non--mesonic decay widths
in theoretical and experimental analyses.
\end{abstract}


\end{frontmatter}


Hypernuclei, bound systems of neutrons, protons and hyperons,
embody an important source of information
able to match nuclear and particle physics. On the one
hand, studies on the production mechanisms and the structure of
hypernuclei \cite{Ha06} are of interest since they provide indications
on the hyperon--nucleon and hyperon--hyperon strong interactions
which cannot be determined with precision from
scattering experiments (such experiments are very challenging due to the
very short hyperon lifetimes). Hypernuclear weak decay is on the other
hand the only available tool to acquire knowledge on the baryon--baryon
strangeness--changing interactions \cite{Ra98,Al02,Pa07,Ou05}.
The above subjects, which presuppose the solution of complex
many--body problems,
are in turn crucially related to the renormalization
of hyperons and mesons properties in the nuclear medium and
are relevant in connection with neutron star studies \cite{Sch08}.

A $\Lambda$--hypernucleus weakly decays via two distinct modes:
the mesonic decay, which concerns the direct disappearance of the hyperon,
$\Lambda \to \pi^- p$ and $\Lambda \to \pi^0 n$,
and the non--mesonic decay, which occurs
through the hyperon interactions with one or more nucleons of the
medium: $\Lambda N\to nN$, $\Lambda NN\to nNN$, etc, where
$N=n$ or $p$. The total non--mesonic decay rate is
indicated by $\Gamma_{\rm NM}=\Gamma_1+\Gamma_2$ in the present study,
$\Gamma_1=\Gamma_n+\Gamma_p$ and $\Gamma_2=\Gamma_{nn}+\Gamma_{np}+
\Gamma_{pp}$ being the one-- and two--nucleon induced widths, respectively.
Moreover, we introduce the definitions:
$\Gamma_n\equiv \Gamma(\Lambda n\to nn)$,
$\Gamma_p\equiv \Gamma(\Lambda p\to np)$,
$\Gamma_{nn}\equiv \Gamma(\Lambda nn\to nnn)$,
$\Gamma_{np}\equiv \Gamma(\Lambda np\to nnp)$ and
$\Gamma_{pp}\equiv \Gamma(\Lambda pp\to npp)$.

The non--mesonic
weak decay of hypernuclei has been studied quite extensively to date both
theoretically \cite{Ra98,Al02,Pa07} and experimentally \cite{Ou05}.
In the latest years a substantial amount of data has been collected
at the KEK laboratory \cite{KEK} and by the FINUDA experiment
at Daphne \cite{FI}. Future measurements will be carried out at J--PARC
\cite{jparc,Aj-jparc} and GSI \cite{GSI}. The development of innovative
experimental techniques and elaborated theoretical models has allowed
to reach a reasonable agreement between data and predictions for the
non--mesonic decay rate $\Gamma_{\rm NM}$,
the $\Gamma_n/\Gamma_p$ ratio and the
intrinsic asymmetry parameter $a_\Lambda$ \cite{Ch08}.
Concerning the most recent developments, we point out the finding that
the inclusion of a two--pion--exchange mechanism
(in both the uncorrelated and correlated fashions) in the weak transition
potential seems to play a crucial role in solving the
asymmetry puzzle \cite{Ch07}. A recent and different approach must be
mentioned which proved that the exchange of the axial--vector
$a_1$--meson is also relevant in asymmetry calculations \cite{It08}.

Despite this progress, one should observe that no experimental
identification has been obtained yet of two--nucleon stimulated decays,
with the exception of a couple of rather indirect results
\cite{Bh07,Par07}. An indirect signature was found in KEK
single and double--coincidence nucleon spectra from non--mesonic
decay by Ref.~\cite{Bh07},
which however was based on a rather simplistic analysis
regarding a supposed phase space argument allowing a uniform
sharing of the decay $Q$--value among the two or three final nucleons
in one-- or two--nucleon induced decays.
Such analysis provided an indication for a ratio
$\Gamma_2/\Gamma_1\simeq 0.7$ for $^{12}_\Lambda$C, while values
scattered in the interval $0.2$-$0.5$ were obtained theoretically
\cite{Al91,Ra94,Al00,ba04,Ba06,ba08}.
Another experiment \cite{Par07}, performed at BNL, obtained an upper limit for
$^4_\Lambda$He, $\Gamma_2/\Gamma_1<0.32$ (95\% CL),
which seems to be incompatible with the indication of Ref.~\cite{Bh07}.
It is thus clear that improved theoretical and experimental
determinations of $\Gamma_2$ are necessary.
An experiment with this purpose is indeed planned at J--PARC \cite{jparc} and
will adopt double-- and triple--coincidence nucleon measurements.
Possibly, such kind of study will enable an improved determination
of the whole set of non--mesonic decay widths. This is an essential
matter for a further development of the field, aimed at achieving
a detailed understanding of the reaction mechanisms underlying the various
weak decay channels.

Various theoretical papers were dedicated to the calculation of the rates
for the two--nucleon stimulated decay \cite{Al91,Ra94,Al00,ba04,Ba06,ba08,Al00b}.
Some works also analyzed the effects of this mechanism
on the observable non--mesonic decay nucleon spectra \cite{Ra97,Al00,Ba06,Ga03}.
The first paper which took into account this decay mode
was Ref.~\cite{Al91}. There, within a nuclear matter scheme
based on the polarization propagator method of Ref.~\cite{os85},
a phenomenological description of the two--particle two--hole
($2p2h$) polarization propagator was introduced
by adapting previous results by the same authors on electron scattering off nuclei
to nuclear pion absorption.
This approach was improved in Ref.~\cite{Ra94} and then in Ref.~\cite{Al00}.
Again, the $2p2h$ configurations were not
calculated exactly in Refs.~\cite{Ra94,Al00}, but a phase space argument
together with data on real pion absorption in nuclei was adopted. Subsequently,
Ref.~\cite{Al00b} evaluated microscopically the non--mesonic weak decay rates
by means of a path integral method which allowed a classification of the
$2p2h$ contributions according to the so--called boson loop expansion.
For technical reasons, it was not possible to
separate the total width $\Gamma_{\rm NM}=\Gamma_1+\Gamma_2$ into
the partial contributions $\Gamma_1$ and $\Gamma_2$ in Ref.~\cite{Al00b}.
We also point out that only the decay channel
$\Lambda np\to nnp$ was considered in the discussions of
Refs.~\cite{Al91,Ra94,Al00,Al00b}. On the contrary,
all the three two--nucleon induced channels were explicitly taken into account
in the microscopic approach of Ref.~\cite{ba04}, which evaluated
the corresponding direct $2p2h$ $\Lambda$ self--energy diagrams induced
by ground state correlations. The adopted weak transition potential
included the exchange of the full pseudoscalar ($\pi$, $\eta$, $K$)
and vector meson ($\rho$, $\omega$, $K^*$) octets while the
residual strong interaction was described by a well--tested
Bonn potential model embodying the exchange of $\pi$, $\rho$, $\sigma$
and $\omega$ mesons. As a result of the distinct
approaches followed in Refs.~\cite{Ra94,ba04}, the
kinematics of the nucleons emitted in two--nucleon induced decays
turned out to be very different in these two studies,
as discussed in Ref.~\cite{Ba06}, resulting in final spectra
which could be well discriminated by a future
triple--nucleon coincidence experiment.

We would also like to stress that a proper determination of the
widths and the nucleon spectra for two--nucleon induced decays is
essential, in turn, for an accurate determination of the
$\Gamma_n/\Gamma_p$ ratio (undoubtedly, analyses of this ratio are
also influenced by nucleon final state interaction effects \cite{Ra97}).
On the one hand, this was demonstrated by
Refs.~\cite{Ga03,Ba06} in theoretical analyses of experimental
double--coincidence nucleon spectra which
allowed to derive values of $\Gamma_n/\Gamma_p$ in agreement
with pure theoretical calculations. The extracted values of $\Gamma_n/\Gamma_p$
turned out to be rather sensitive to the input used for $\Gamma_2$.
On the other hand, among the various non--mesonic decay rates,
only $\Gamma_{\rm NM}=\Gamma_1+\Gamma_2$ is directly
accessible to experiments, where it is derived from measurements
of the total lifetime and the mesonic rate as
$\Gamma_{\rm NM}=\Gamma_{\rm T}-\Gamma_{\rm M}$. Therefore, again one sees
that the decay rate $\Gamma_2$ plays an essential role in the determination
of $\Gamma_n/\Gamma_p$: only after a genuine disentanglement between
$\Lambda N\to nN$ and $\Lambda NN\to nNN$ experimental events one can
proceed to an analysis of the nucleon spectra and deduce the experimental
value of $\Gamma_n/\Gamma_p$.

In the present Letter the effect of Pauli exchange terms in the two--nucleon
induced hypernuclear decay is investigated for the first time.
We shall apply the calculation to the hypernucleus $^{12}_\Lambda$C.
Exchange corrections are genuine quantum mechanical
effects due to the underlying Fermi--Dirac statistics which
are expected to be important in the present case as for any fermionic
many--body system. Exchange terms, required by fermion antisymmetrization,
turned out to be relevant in one--nucleon induced decays,
both in tree level \cite{Ji01,ba03} and in final state interaction
(RPA--like) diagrams \cite{Ba07}.
We employ the non--relativistic nuclear matter formalism extended to
finite nuclei by the local density approximation which was established
in Refs.~\cite{ba03,ba04}. The weak transition potential,
whose formulation and weak coupling constants are
taken from Refs.~\cite{Pa97},
contains the exchange of the mesons of the pseudoscalar
and vector octets, $\pi$, $\eta$, $K$, $\rho$, $\omega$ and $K^*$.
The strong coupling constants and cut--off parameters entering the
weak transition potential are instead
deduced from the Nijmegen soft--core interaction NSC97f of Ref.~\cite{st99}.
For the nucleon--nucleon strong interaction entering the $2p2h$ correlations
we adopt, as in Ref.~\cite{ba04}, a Bonn potential with the exchange
of $\pi$, $\rho$, $\sigma$ and $\omega$ mesons \cite{ma87,br96}.
All the two--nucleon stimulated channels, $\Lambda nn\to nnn$, $\Lambda np\to nnp$
and $\Lambda pp\to npp$, are included in the calculation.


For the formal derivation of the one--nucleon induced decay widths,
which includes both direct and exchange contributions,
we refer to the original discussion of Ref.~\cite{ba03}.

Consider then the two--nucleon induced decay width
for a $\Lambda$--hyperon with four--momentum $k=(k_0,\v{k})$
inside infinite nuclear matter with Fermi momentum
$k_F$. Let us write it in a schematic way as follows:
\begin{equation}
\label{decw} \Gamma_{2}(k,k_{F}) = \sum_{f} \,
 |\bra{f} V^{\Lambda N\to NN} \ket{0}_{k_{F}}|^{2}
\delta (E_{f}-E_{0})~,
\end{equation}
$\ket{0}_{k_{F}}$ and $\ket{f}$ denoting, respectively, the initial
hypernuclear ground state (whose energy is $E_0$)
and the possible final $3p2h$ states (with energy $E_f$) corresponding
to three--nucleon emission. Moreover, $V^{\Lambda N\to NN}$ is the weak
transition potential.

The two--nucleon induced decay rate for a finite hypernucleus is obtained
by the local density approximation~\cite{os85},
i.e., after averaging the above partial width over the $\Lambda$
momentum distribution in the considered hypernucleus,
$|\widetilde{\psi}_{\Lambda}(\v{k})|^2$, and over the local Fermi momentum,
$k_{F}(r) = \{3 \pi^{2} \rho(r)/2\}^{1/3}$,
$\rho(r)$ being the density profile of the hypernuclear core.
One thus has:
\begin{equation}
\label{decwpar3}
\Gamma_{2} = \int d \v{k} \, |\widetilde{\psi}_{\Lambda}(\v{k})|^2
\int d \v{r} \, |\psi_{\Lambda}(\v{r})|^2
\Gamma_2(\v{k},k_{F}(r))~,
\end{equation}
where for $\psi_{\Lambda}(\v{r})$, the Fourier transform of
$\widetilde{\psi}_{\Lambda}(\v{k})$, we
adopt the $1s_{1/2}$ harmonic oscillator wave--function with
frequency $\hbar \omega=10.8$ MeV adjusted to the experimental energy
separation between the $s$ and $p$ $\Lambda$--levels in $^{12}_\Lambda$C.
The $\Lambda$ hyperon total energy in Eqs.~(\ref{decw}) and (\ref{decwpar3})
is given by $k_{0}=m_\Lambda+\v{k}^2/(2 m_\Lambda)+V_{\Lambda}$,
i.e., it also contains an experimental binding term $V_\Lambda=-10.8$ MeV.

\begin{figure}[h]
\centerline{\includegraphics[scale=0.67]{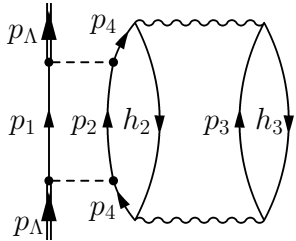}}
\caption{Direct Goldstone diagram
contributing to the two--nucleon induced decay width.}
\label{dir}
\end{figure}

The final states in Eq.~(\ref{decw}) are restricted to
three--particle emission. Since
$V^{\Lambda N\to NN}$ is a two--body operator, two--nucleon
induced decays originates from ground state correlations
due to the nucleon--nucleon interaction.
The normalized hypernuclear ground state wave--function
can be written as \cite{ba08}:
\begin{equation}
\label{gstate} \ket{0}_{k_{F}}=\mathcal{N}(k_{F}) \, \left(\ket{\;}
- \sum_{p_{4}, p_{3}, h_{2}, h_{3}} \,
\frac{\bra{p_{4} p_{3} h_{2} h_{3}} V^{N N} \ket{\;}_{D+E}}
{\varepsilon_{p4}+\varepsilon_{p3}-\varepsilon_{h2}-\varepsilon_{h3}} \,
\ket{p_{4} p_{3} h_{2} h_{3}}\right)~,
\end{equation}
where $\ket{\;}$ is the uncorrelated ground state wave--function,
i.e., the Hartree--Fock vacuum, while the second term in the rhs
represents $2p2h$ correlations and contains both direct ($D$) and exchange
($E$) matrix elements of the residual strong interaction $V^{N N}$.
Besides, the particle and hole energies are denoted by $\varepsilon_{i}$
and $\mathcal{N}(k_{F})$ is the normalization factor
\begin{equation}
\label{norconst}
\mathcal{N}(k_{F})=\left( 1 + \sum_{p_{4},
p_{3}, h_{2}, h_{3}} \, \left|\frac{\bra{p_{4} p_{3} h_{2} h_{3}}
V^{N N} \ket{\;}_{D+E}}
{\varepsilon_{p4}+\varepsilon_{p3}-\varepsilon_{h2}-\varepsilon_{h3}}
\right|^{2}
\, \right)^{-1/2}~.
\end{equation}
The particular labeling of Eqs.~(\ref{gstate}) and (\ref{norconst})
is easily understood
from the direct $2p2h$ $\Lambda$ self--energy diagram of Fig.~\ref{dir}.

Inserting Eq.~(\ref{gstate}) into Eq.~(\ref{decw}) one obtains:
\begin{eqnarray}
\label{decw2}
\Gamma_{2}(\v{k},k_{F}) & = & \mathcal{N}^{2}(k_{F})  \sum_{p_{1},
p_{2}, p_{3}, p_{4}, h_{2}, h_{3}} \,
\Biggl|\bra{p_{1} p_{2} p_{3} h_{2} h_{3}} V^{\Lambda N\to NN}
\ket{p_{\Lambda} p_{4} p_{3} h_{2} h_{3}}_{D'+E'} \nonumber \\
&& \times
\frac{\bra{p_{\Lambda} p_{4} p_{3} h_{2} h_{3}} V^{N N} \ket{p_{\Lambda}}_{D+E}}
{\varepsilon_{p4}+\varepsilon_{p3}-\varepsilon_{h2}-\varepsilon_{h3}} \Biggr|^{2}
\delta (E_{\ket{p_{1} p_{2} p_{3} h_{2} h_{3}}}-E_{0})~, \nonumber
\end{eqnarray}
where also the matrix elements of the weak transition potential
$V^{\Lambda N\to NN}$ appears in the antisymmetrized form.
In the $V^{N N}$ matrix elements the $\Lambda$ is acting as a spectator
and so the particle $p_3$ and the
holes $h_{2}, \, h_{3}$ in the $V^{\Lambda N\to NN}$ matrix elements.
In order to lighten the notation, we omit the spectators and write:
\begin{eqnarray}
\label{decw3}
&&\Gamma_{2}(\v{k},k_{F})  = \mathcal{N}^{2}(k_{F}) \nonumber \\
&&\times \sum_{p_{1}, p_{2}, p_{3}, p_{4}, h_{2}, h_{3}} \,
\Biggl|\left[ \bra{p_{1} p_{2}} V^{\Lambda N\to NN} \ket{p_{\Lambda} p_{4}}_{D'} +
\bra{p_{1} p_{2}} V^{\Lambda N\to NN} \ket{p_{\Lambda} p_{4}}_{E'} \right]
\nonumber \\
&& \times
\left[\frac{\bra{p_{4} p_{3}} V^{N N} \ket{h_{2} h_{3}}_{D}}{\epsilon_{2p2h}} +
\frac{\bra{p_{4} p_{3}} V^{N N} \ket{h_{2} h_{3}}_{E}}{\epsilon'_{2p2h}}
\right] \Biggr|^{2} \delta (E_{\ket{p_{1} p_{2} p_{3} h_{2} h_{3}}}-E_{0})~, \nonumber
\end{eqnarray}
where $\epsilon_{2p2h}$ and $\epsilon'_{2p2h}$ are the energy denominators which
correspond to the direct and exchange matrix elements.
In order to make more evident the set of terms contributing to the
two--nucleon induced rate, it is convenient
to expand the above expression as follows:
\begin{eqnarray}
\label{decw4}
&&\Gamma_{2}(\v{k},k_{F}) =  \mathcal{N}^{2}(k_{F})   \\
&& \times \sum_{p_{1}, p_{2}, p_{3}, p_{4}, h_{2}, h_{3}} \,
\left[\frac{\bra{ h_{2} h_{3}} (V^{N N})^{\dagger} \ket{p_{4} p_{3}}_{D}}{\epsilon_{2p2h}} +
\frac{\bra{h_{2} h_{3}} (V^{N N})^{\dagger} \ket{p_{4} p_{3}}_{E}}{\epsilon'_{2p2h}}\right]
\nonumber \\
&& \times
\left[\bra{p_{\Lambda} p_{4}} (V^{\Lambda N\to NN})^{\dagger} \ket{p_{1} p_{2}}_{D'} +
\bra{p_{\Lambda} p_{4}} (V^{\Lambda N\to NN})^{\dagger} \ket{p_{1} p_{2}}_{E'}\right]
\nonumber \\
&& \times
\left[\bra{p_{1} p_{2}} V^{\Lambda N\to NN} \ket{p_{\Lambda} p_{4}}_{D'} +
\bra{p_{1} p_{2}} V^{\Lambda N\to NN} \ket{p_{\Lambda} p_{4}}_{E'}\right]
\nonumber \\
&& \times
\left[\frac{\bra{p_{4} p_{3}} V^{N N} \ket{ h_{2} h_{3}}_{D}}{\epsilon_{2p2h}} +
\frac{\bra{p_{4} p_{3}} V^{N N} \ket{ h_{2} h_{3}}_{E}}{\epsilon'_{2p2h}} \right]
\delta (E_{\ket{p_{1} p_{2} p_{3} h_{2} h_{3}}}-E_{0})~. \nonumber
\end{eqnarray}

\begin{figure}[h]
\centerline{\includegraphics[scale=0.63]{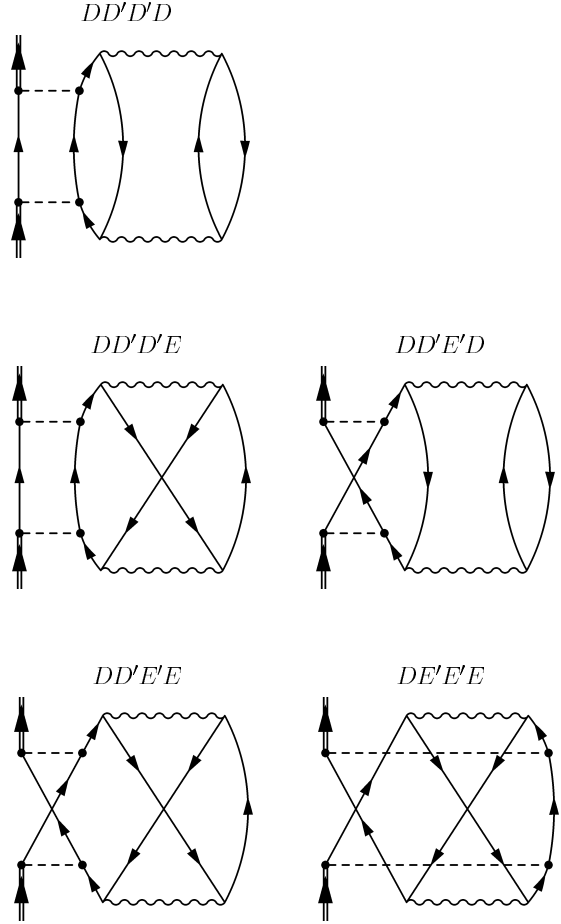}}
\caption{Direct and exchange Goldstone diagrams induced by
ground state correlations and contributing to the
two--nucleon induced decay width of Eqs.~(\ref{prodde}) and
(\ref{totald+e}).}
\label{direxc}
\end{figure}

The product of the four direct plus exchange pieces makes a total of sixteen
contributions to the decay rate of Eq.~(\ref{decw4}).
We identify each one of them by the notation $\Gamma_2^{PQ'R'S}$,
where $P,\, Q,\, R,\, S=D$ (direct) or $E$ (exchange).
For instance, $P$ refers to the direct or exchange
character of the matrix element
$\bra{ h_{2} h_{3}} (V^{N N})^{\dagger} \ket{p_{4} p_{3}}_P$
(the meaning of $Q, \, R$ and $S$ is thus self--evident).
Several of these contributions turn out to be equal among each other (for instance,
$\Gamma_2^{DD'D'D}=\Gamma_2^{DE'E'D}=\Gamma_2^{ED'D'E}=\Gamma_2^{EE'E'E}$) and there are a
total of five different corresponding self--energy diagrams. They are depicted in
Fig.~\ref{direxc} and amount to the following partial rates:
\begin{eqnarray}
\label{prodde}
\Gamma_2^{dd'd'd} & \equiv & \frac{1}{4}
(\Gamma_2^{DD'D'D}+\Gamma_2^{DE'E'D}+\Gamma_2^{ED'D'E}+\Gamma_2^{EE'E'E})=\Gamma_2^{DD'D'D}~, \\
\Gamma_2^{dd'd'e} & \equiv & \frac{1}{4}
(\Gamma_2^{DD'D'E}+\Gamma_2^{ED'D'D}) = \frac{1}{2}\Gamma_2^{DD'D'E}~,
\nonumber \\
\Gamma_2^{dd'e'd} & \equiv & \frac{1}{4}
(\Gamma_2^{DD'E'D}+\Gamma_2^{DE'D'D}+\Gamma_2^{ED'E'E}+\Gamma_2^{EE'D'E})=\Gamma_2^{DD'E'D}~,
\nonumber \\
\Gamma_2^{dd'e'e} & \equiv & \frac{1}{4}
(\Gamma_2^{DD'E'E}+\Gamma_2^{DE'D'E}+\Gamma_2^{ED'E'D}+\Gamma_2^{EE'D'D})=\Gamma_2^{DD'E'E}~,
\nonumber \\
\Gamma_2^{de'e'e} & \equiv & \frac{1}{4}
(\Gamma_2^{DE'E'E}+\Gamma_2^{EE'E'D}) = \frac{1}{2}\Gamma_2^{DE'E'E}~. \nonumber
\end{eqnarray}

Details on the calculation of the partial decay rates of Eq.~(\ref{prodde})
will be given elsewhere.

The two--nucleon stimulated decay rate of Eqs.~(\ref{decwpar3}) and
(\ref{decw4}) is therefore obtained as:
\begin{equation}
\label{totald+e}
\Gamma^{\rm pp}_2=
\Gamma_2^{dd'd'd} + \Gamma_2^{dd'd'e} + \Gamma_2^{dd'e'd}
+ \Gamma_2^{dd'e'e} + \Gamma_2^{de'e'e}~, \nonumber
\end{equation}
where the use of the $\rm pp$ apex is due to the fact that
in the previous derivation we have limited ourself to self--energy
diagrams (given in Figure~\ref{direxc})
with the two weak transition potentials
connected to the same particle line.
By denoting with $\Gamma_2^{{\rm pp}\, D}$ and $\Gamma_2^{{\rm pp}\, E}$
the direct and exchange decay rates corresponding to the result
of Eq.~(\ref{totald+e}), it follows that $\Gamma_2^{\rm pp}
=\Gamma_2^{{\rm pp}\, D}+\Gamma_2^{{\rm pp}\, E}$,
with $\Gamma_2^{{\rm pp}\, D}=\Gamma_2^{dd'd'd}$ and
$\Gamma_2^{{\rm pp}\, E}= \Gamma_2^{dd'd'e} + \Gamma_2^{dd'e'd}
+ \Gamma_2^{dd'e'e} + \Gamma_2^{de'e'e}$.

\begin{figure}[h]
\centerline{\includegraphics[scale=0.67]{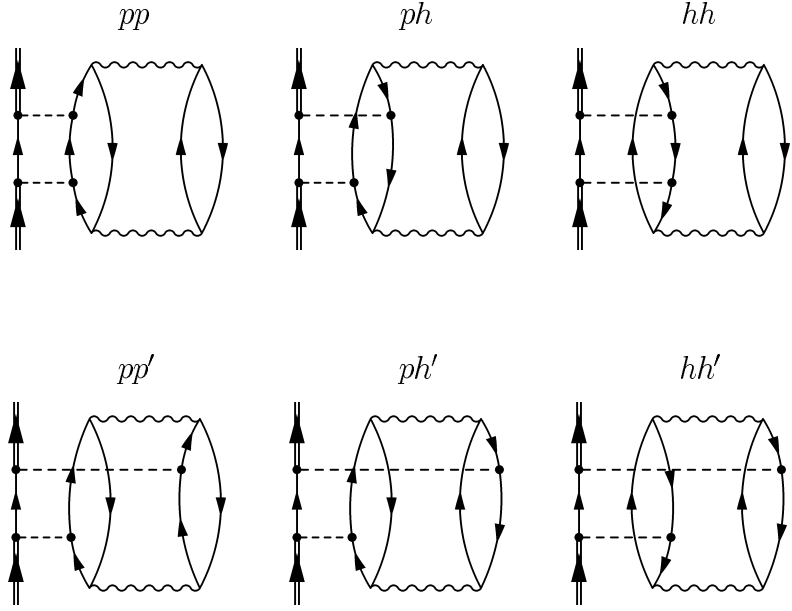}}
\caption{Direct Goldstone diagrams induced by ground state correlations and
contributing to the two--nucleon induced decay width of Eq.~(\ref{totalg2}).}
\label{alldirect}
\end{figure}

At this point we have to note that
also diagrams with at least one of the two
weak transition potentials connected to a hole line
($\Gamma_2^{{\rm ph}\, P}$ and $\Gamma_2^{{\rm hh}\, P}$ contributions,
with $P=D$ or $E$) are expected to intervene
(the direct Goldstone diagrams corresponding to
$\Gamma_2^{{\rm pp}\, D}$, $\Gamma_2^{{\rm ph}\, D}$ and
$\Gamma_2^{{\rm hh}\, D}$ are displayed in the upper part
of Figure~\ref{alldirect}).
Moreover, diagrams with the weak transition potentials connected
to two different particle--hole bubbles must be taken into account as well,
resulting in the contributions $\Gamma_2^{{\rm pp'}\, P}$, $\Gamma_2^{{\rm ph'}\, P}$
and $\Gamma_2^{{\rm hh'}\, P}$
(the Goldstone diagrams corresponding to the direct parts are given
in the lower part of Figure~\ref{alldirect}).
Once the whole set of
$2p2h$ diagrams induced by ground state correlations is considered,
the total two--nucleon induced decay width is obtained as:
\begin{eqnarray}
\label{totalg2}
\Gamma_2&=&
\Gamma^{{\rm pp}}_2+\Gamma^{{\rm ph}}_2+\Gamma^{{\rm hh}}_2
+\Gamma^{{\rm pp'}}_2 +\Gamma^{{\rm ph'}}_2+\Gamma^{{\rm hh'}}_2\\
&=&\Gamma^{{\rm pp}\, D}_2+\Gamma^{{\rm pp}\, E}_2
+\Gamma^{{\rm ph}\, D}_2+\Gamma^{{\rm ph}\, E}_2
+\Gamma^{{\rm hh}\, D}_2+\Gamma^{{\rm hh}\, E}_2 \nonumber \\
&&+\Gamma^{{\rm pp'}\, D}_2+\Gamma^{{\rm pp'}\, E}_2
+\Gamma^{{\rm ph'}\, D}_2+\Gamma^{{\rm ph'}\, E}_2
+\Gamma^{{\rm hh'}\, D}_2+\Gamma^{{\rm hh'}\, E}_2~. \nonumber
\end{eqnarray}
The three isospin channels, $\Lambda nn\to nnn$, $\Lambda np\to nnp$ and
$\Lambda pp\to npp$, contribute to each term $\Gamma^{{\rm ij}\, P}_2$:
\begin{equation}
\label{partiisospin}
\Gamma^{{\rm ij}\, P}_2=
\Gamma^{{\rm ij}\, P}_{nn}+\Gamma^{{\rm ij}\, P}_{np}+\Gamma^{{\rm ij}\, P}_{pp}
\,\,\,\,\,\,\,\,(P=D\, {\rm or}\, E)~,
\end{equation}
and, in turn, each $\Gamma^{{\rm ij}\, D}_{N_1N_2}$ ($\Gamma^{{\rm ij}\, E}_{N_1N_2}$)
is obtained from four (twelve) direct (exchange) diagrams.


In Table~\ref{gamm2ant} we give our results for the partial rates
$\Gamma^{pq'r's}_2$ of Eq.~(\ref{prodde}) and (\ref{totald+e})
for the case of $^{12}_{\Lambda}C$. We emphasize that
these predictions have been obtained by an exact calculation from the
Goldstone rules applied to the diagrams of Figure~\ref{direxc}.
In the first three lines we supply the separate
contributions to the three isospin channels;
the summed results are listed in the last line.
As expected, the dominant contribution
is provided by the direct term $\Gamma^{dd'd'd}_2$, while among the
exchange terms the more important ones turn out to be $\Gamma^{dd'd'e}_2$
and $\Gamma^{dd'ed}_2$. Their negative sign is due to the odd number of
crossing between fermionic lines.
The global effect of antisymmetry is to decrease by 32\% the
two--nucleon induced rate
$\Gamma^{\rm pp}_2=\Gamma^{\rm pp}_{nn}+\Gamma^{\rm pp}_{np}+\Gamma^{\rm pp}_{pp}$
obtained with the direct terms only.
The $\Lambda np\to nnp$ channel remains the most important one,
in agreement with a quasi--deuteron approximation, and
$\Gamma^{\rm pp}_{np}:\Gamma^{\rm pp}_{pp}:\Gamma^{\rm pp}_{nn}=
0.78:0.16:0.06$. This relation remains almost unaltered when only the direct
terms are taken into account.

\begin{table}[h]
\begin{center}
\caption{Partial contributions to the two--nucleon induced decay
width of Eq.~(\ref{totald+e}) for $^{12}_{\Lambda}C$ in units of the free
$\Lambda$ rate, $\Gamma^0= 2.52 \cdot 10^{-6}$ eV.
The first column indicates the three different
isospin channels contributing to $\Gamma^{\rm pp}_{2}$.
Values smaller than $0.0005$ are represented by $\sim 0$.
}
\label{gamm2ant}
\begin{tabular}{ccccccc}   \hline\hline
Channel & ~~${\Gamma}^{dd'd'd}_{2}$~~ & ~~${\Gamma}^{dd'd'e}_{2}$~~ & ~~${\Gamma}^{dd'e'd}_{2}$~~
& ~~${\Gamma}^{dd'e'e}_{2}$~~ & ~~${\Gamma}^{de'e'e}_{2}$~~ & ~~${\Gamma}^{\rm pp}_{2}$~~
\\  \hline
$\Lambda nn\to nnn$ & $0.012$  & $-0.002$  & $-0.002$  &$\sim 0$ & $0.001$  & $0.009$   \\
$\Lambda np\to nnp$ & $0.184$  & $-0.047$  & $-0.039$  & $0.009$ & $0.016$  & $0.123$   \\
$\Lambda pp\to npp$ & $0.036$  & $-0.008$  & $-0.007$  & $0.002$ & $0.002$  & $0.025$   \\ \hline
sum                 & $0.232$  & $-0.057$  & $-0.048$  & $0.011$ & $0.019$  & $0.157$   \\ \hline\hline
\end{tabular}
\end{center}
\end{table}


We now proceed to discuss
our calculation of the full set of two--nucleon induced contributions
of Eqs.~(\ref{totalg2}) and (\ref{partiisospin}).
All the direct terms have been obtained from the corresponding
Goldstone diagrams of Figure~\ref{alldirect}.
We remind the reader that the direct terms
${\Gamma}^{{\rm pp}\, D}_{2}$, ${\Gamma}^{{\rm ph}\, D}_{2}$
and ${\Gamma}^{{\rm hh}\, D}_{2}$
(${\Gamma}^{{\rm pp'}\, D}_{2}$, ${\Gamma}^{{\rm ph'}\, D}_{2}$
and ${\Gamma}^{{\rm hh'}\, D}_{2}$) were calculated
for the first time in Ref.~\cite{ba04} (Ref.~\cite{ba08}).
Contrarily, we have not evaluated exactly the exchange
terms ${\Gamma}^{{\rm ph}\, E}_{2}$,
${\Gamma}^{{\rm hh}\, E}_{2}$, ${\Gamma}^{{\rm pp'}\, E}_{2}$,
${\Gamma}^{{\rm ph'}\, E}_{2}$ and ${\Gamma}^{{\rm hh'}\, E}_{2}$.
We note that these contributions
are expected to be considerably smaller than the one we calculated exactly,
${\Gamma}^{{\rm pp}\, E}_{2}$. Moreover, each ${\Gamma}^{{\rm ij}\, E}_{2}$ is
smaller in absolute value than ${\Gamma}^{{\rm ij}\, D}_{2}$.
Therefore, we anticipate a limited effect of all the
exchange contributions but ${\Gamma}^{{\rm pp}\, E}_{2}$.

Anyhow, we have evaluated all the exchange terms apart from ${\Gamma}^{{\rm pp}\, E}_{2}$
in an approximated way through the following strategy.
We introduce a Landau--Migdal model in which
the residual strong interaction $V^{NN}$ and the
weak transition potential $V^{\Lambda N\to NN}$ are modified by the addition of
spin--isospin $g'$ and $g'_\Lambda$ parameters:
$V^{NN}(q)\to V^{NN}(q)+(f_\pi/m_\pi)^2 g' \, F^2(q)\,
\v{\sigma} \cdot \v{\sigma}^\prime \, \v{\tau} \cdot \v{\tau}^\prime$ and
$V^{\Lambda N\to NN}(q)\to V^{\Lambda N\to NN}(q)+G_F m^2_\pi
(f_\pi/m_\pi)(B_\pi/2\bar M) g'_\Lambda \,
F^2(q)\, \v{\sigma} \cdot \v{\sigma}^\prime \, \v{\tau} \cdot \v{\tau}^\prime$.
In these expressions, a form factor $F(q)=(\Lambda^2-m^2_\pi)/(\Lambda^2-q_0^2+\v{q}^2)$
with cut--off $\Lambda=1.75$ GeV is included, while $G_F$ is the Fermi constant,
$B_\pi$ is the constant which defines the parity--conserving
$\Lambda \pi N$ weak vertex and
$\bar M$ is the average between the nucleon and $\Lambda$ masses.
The values of $g'$ and $g'_\Lambda$ are fixed by using the results we obtained
microscopically for the ${\Gamma}^{pq'r's}_{2}$'s entering Eq.~(\ref{totald+e}).
The direct term ${\Gamma}^{dd'd'd}_{2}$ has to be calculated both
with and without the modified $V^{NN}$ and $V^{\Lambda N\to NN}$ interactions.
The condition that the calculation with the modified $V^{NN}$
($V^{\Lambda N\to NN}$) provide the same
result of the sum ${\Gamma}^{dd'd'd}_{2}+{\Gamma}^{dd'd'e}_{2}$
(${\Gamma}^{dd'd'd}_{2}+{\Gamma}^{dd'e'd}_{2}$)
computed with the original interactions allows us to have a first
determination of the Landau--Migdal parameter $g'$ ($g'_\Lambda$).
By maintaining unaltered their ratio, the values of the two parameters are
then fine--tuned in order to give the same result of the sum
${\Gamma}^{dd'd'd}_{2}+{\Gamma}^{dd'd'e}_{2}+{\Gamma}^{dd'e'd}_{2}
+{\Gamma}^{dd'e'e}_{2}+{\Gamma}^{de'e'e}_{2}$
when implemented together in the direct term ${\Gamma}^{dd'd'd}_{2}$.
The so determined $g'$ and $g'_\Lambda$
are then used to evaluate the other exchange terms via the relation:
\begin{equation}
{\Gamma}^{{\rm ij}\, E}_{2}={\Gamma}^{{\rm ij}\, D}_{2}(g',g'_\Lambda)
-{\Gamma}^{{\rm ij}\, D}_{2}(g'=g'_\Lambda=0)~\,\,\,\,\,\,\,\,
(ij\neq {\rm pp})~.
\end{equation}

We stress that, formally, the above computational method
makes use of the same Landau--Migdal phenomenology
which have been widely adopted in many applications to take
care of a number of effects such as the baryon--baryon short range correlations
and the LLEE effect. Here we have applied this model only to obtain
approximate results for those exchange diagrams that we have not evaluated
microscopically. This fact is explained by the smallness of the parameters
that we have obtained: $g'=0.05$ and $g'_\Lambda=0.09$.
We also point out that the above mentioned fine--tuning has required
a modification of $g'$ and $g'_\Lambda$ by only 10\%, thus
demonstrating a certain reliability of our approximate procedure.

In Table~\ref{gamma2complete} we report our predictions for the
two--nucleon induced terms $\Gamma^{{\rm ij}\, P}_{N_1N_2}$ entering
Eqs.~(\ref{totalg2}) and (\ref{partiisospin}).
For the three isospin channels the dominant contribution is given by
$\Gamma^{{\rm pp}\, D}_{N_1N_2}$. The next contributions
in order of importance are $\Gamma^{{\rm pp'}\, D}_{N_1N_2}$ and then
the exchange terms that we have
calculated exactly, i.e., $\Gamma^{{\rm pp}\, E}_{N_1N_2}$.
Among the exchange contributions
that we have evaluated approximately, the most important ones
are $\Gamma^{{\rm pp'}\, E}_{N_1N_2}$.
Each exchange term is smaller in absolute
value than the corresponding direct term and
$|\Gamma^{{\rm ij}\, E}_{N_1N_2}/\Gamma^{{\rm ij}\, D}_{N_1N_2}|=0.2$-$0.5$.
By neglecting the exchange terms that we have calculated with the approximated
procedure, the total two--nucleon induced decay width would be $\Gamma_2=0.287$
instead of $0.252$. This comparison shows that the exact calculation
of all the exchange terms may be important for a precise determination of $\Gamma_2$.
From the final results of the last column of Table~\ref{gamma2complete} we see that
$\Gamma_{np}:\Gamma_{pp}:\Gamma_{nn}=0.83:0.12:0.04$, a relation which remains
practically unchanged when limiting to the direct part only.
\begin{table}[h]
\begin{center}
\caption{Partial contributions to the two--nucleon induced decay
width of Eqs.~(\ref{totalg2}) and (\ref{partiisospin})
for $^{12}_{\Lambda}C$ in units of the free $\Lambda$ rate.
Values smaller than $0.0005$ are represented by $\sim 0$.}
\label{gamma2complete}
\resizebox*{\textwidth}{!}{
\begin{tabular}{cccccccc}   \hline\hline
Channel &
~~${\Gamma}^{{\rm pp}\, D}_{2}$~~ &
~~${\Gamma}^{{\rm pp}\, E}_{2}$~~ &
~~${\Gamma}^{{\rm ph}\, D}_{2}$~~ &
~~${\Gamma}^{{\rm ph}\, E}_{2}$~~ &
~~${\Gamma}^{{\rm hh}\, D}_{2}$~~ &
~~${\Gamma}^{{\rm hh}\, E}_{2}$~~ &
                                  \\ \hline
$\Lambda nn\to nnn$ & $0.012$ & $-0.003$ & $\sim 0$  &  $\sim 0$   & $0.002$  & $-0.001$   &    \\
$\Lambda np\to nnp$ & $0.184$ & $-0.061$ & $-0.003$   &   $0.001$    & $0.027$  & $-0.006$   &    \\
$\Lambda pp\to npp$ & $0.036$ & $-0.011$ & $\sim 0$  &  $\sim 0$   & $0.006$  & $-0.001$   &    \\ \hline
sum                 & $0.232$ & $-0.075$ & $-0.003$   &   $0.001$    & $0.035$  & $-0.008$   &    \\ \hline\hline
Channel &
~~${\Gamma}^{{\rm pp'}\, D}_{2}$~~ &
~~${\Gamma}^{{\rm pp'}\, E}_{2}$~~ &
~~${\Gamma}^{{\rm ph'}\, D}_{2}$~~ &
~~${\Gamma}^{{\rm ph'}\, E}_{2}$~~ &
~~${\Gamma}^{{\rm hh'}\, D}_{2}$~~ &
~~${\Gamma}^{{\rm hh'}\, E}_{2}$~~ &
~~${\Gamma}_{2}$~~\\ \hline
$\Lambda nn\to nnn$ & $0.002$  & $-0.001$    & $\sim 0$   & $\sim 0$  & $\sim 0$  & $\sim 0$  & $0.011$  \\
$\Lambda np\to nnp$ & $0.081$  & $-0.021$    &  $0.010 $   & $-0.004$   & $0.004$    & $-0.002$   & $0.210$  \\
$\Lambda pp\to npp$ & $0.001$  & $ \sim 0$  & $\sim 0$   & $\sim 0$  & $\sim 0$  & $\sim 0$  & $0.031$ \\ \hline
sum                 & $0.084$  & $-0.022$    &  $0.010 $   & $-0.004$   & $0.004$    & $-0.002$   & $0.252$ \\ \hline\hline
\end{tabular}
}
\end{center}
\end{table}

Our predictions for the total one-- and two--nucleon induced rates
are reported in Table~\ref{gam12} for both the full (direct plus exchange)
calculation and for the evaluation limited to the direct part only.
Note that in the Direct Only calculation the normalization of the ground state
wave--function of Eq.~(\ref{norconst}) only contains
direct matrix elements, while in the evaluation of the direct contributions of
Table~\ref{gamma2complete} (where also exchange contributions are given)
the ground state incorporates both direct and exchange matrix elements.
We see that the antisymmetrization property implies a \emph{reduction} of 18\%
in the decay rate $\Gamma_2$. This is a rather important effect, especially when
combined with the finding that the introduction of
exchange terms produces an \emph{increase} of 19\%
in the value of $\Gamma_1$. We thus find that antisymmetrization entails a substantial
decrease, of 31\%, in the value of $\Gamma_2/\Gamma_1$, while having less
influence on the total non--mesonic decay width $\Gamma_{\rm NM}$ and the
$\Gamma_n/\Gamma_p$ ratio. The complete result for $\Gamma_{\rm NM}$
is in agreement with the most recent KEK datum only
within $1.7\, \sigma$ deviations, while it agrees well with the older KEK experiment.
Further theoretical and experimental analyses are thus necessary to improve
the agreement between predictions and data.
A forthcoming experiment planed at J--PARC \cite{jparc} will
measure the ratio $\Gamma_2/\Gamma_1$ for the first time.
In addition, new data on $\Gamma_{\rm NM}$
will also help the comparison between theory and experiment
(the data in Table~\ref{gam12} are indeed only barely compatible between each other).
Before concluding we note that our result for the ratio between the neutron-- and the
proton--induced rates, $\Gamma_n/\Gamma_p=0.327$, is in agreement
with the value $(\Gamma_n/\Gamma_p)^{\rm Exp}=0.29\pm0.14$ deduced in Ref.~\cite{Ga03}
by fitting KEK--E508 double--coincidence nucleon spectra.
We also point out that the rate $\Gamma_{n}$ is predicted to be smaller
than $\Gamma_{np}$: $\Gamma_{n}=0.154$, $\Gamma_{np}=0.210$.
As a final comment, we mention that
from the notable weight obtained for the exchange contributions
we anticipate important modifications in the measurable nucleon spectra,
which indeed depend on the kinematics of the intermediate nucleons
in the $\Lambda$ self--energy diagrams (we expect such kinematics to be
strongly influenced by fermion antisymmetrization).
\begin{table}[h]
\begin{center}
\caption{One-- and two--nucleon stimulated
non--mesonic weak decay widths of $^{12}_{\Lambda}C$
in units of the free $\Lambda$ rate.
The most recent experimental results
are given for the total non--mesonic rate.
}
\label{gam12}
\begin{tabular}{cccccc}   \hline\hline
 & ${\Gamma}_{1}$ &
 ${\Gamma}_{2}$& ${\Gamma}_{\rm NM}$ & ${\Gamma}_{n}/{\Gamma}_{p}$
 & ${\Gamma}_{2}/{\Gamma}_{1}$
\\  \hline
Direct Only         &  0.524  & 0.308 & 0.832 & 0.310 & 0.588 \\
Direct $+$ Exchange &  0.624  & 0.252 & 0.876 & 0.327 & 0.404 \\ \hline
$\mbox{KEK--E508} \,$ \cite{Ou08}&   &    &     $0.929 \pm 0.027 \pm 0.016$  & &\\
$\mbox{KEK--E307} \,$ \cite{sa05}&   &    &     $0.828 \pm 0.056 \pm 0.066$ & &\\
\hline\hline
\end{tabular}
\end{center}
\end{table}





In this Letter
we have studied the effects of Pauli exchange terms in
the two--nucleon induced weak decay of $^{12}_\Lambda$C hypernuclei.
All the possible stimulating channels, $\Lambda nn\to nnn$,
$\Lambda np\to nnp$ and $\Lambda pp\to npp$, have been considered.
A non--relativistic nuclear matter scheme has been adopted together
with a local density approximation. The employed weak transition
potential (residual strong interaction, modeled on a Bonn potential)
contains the exchange of the full set of mesons of the
pseudoscalar and vector octets ($\pi$, $\rho$, $\sigma$ and $\omega$ mesons).
We predict that $\Gamma_p:\Gamma_{np}:\Gamma_n:\Gamma_{pp}:\Gamma_{nn}=
0.54:0.24:0.18:0.04:0.01$.
The exchange contributions turns out to reduce by 18\%
the value of the two--nucleon induced rate obtained without fermion antisymmetrization.
The global effect of antisymmetrization is of
reducing $\Gamma_2/\Gamma_1$ by 31\% and increasing
$\Gamma_{\rm NM}=\Gamma_1+\Gamma_2$ and $\Gamma_n/\Gamma_p$ by 5\%, while
maintaining practically unchanged the
$\Gamma_{np}:\Gamma_{pp}:\Gamma_{nn}$ relation.
A careful evaluation of the exchange contributions is thus of special
relevance for a correct separation of the total non--mesonic decay rate
into the one-- and two--nucleon induced parts.

In order to achieve a more exhaustive understanding
of non--mesonic weak decay, for the future
we plan to extend the microscopic calculation of the two--nucleon induced
decay modes to the whole set of Pauli exchange diagrams.
Another consequential prosecution of such a study will concern
the calculation of those nucleon spectra which are essential
for a meaningful comparison with experiment as well as
for extracting, from data, the whole set of partial
non--mesonic decay widths. From the results
presented here we can anticipate a pronounced modification, induced
by antisymmetrization, in these measurable nucleon spectra.
Such a study will also involve the use of a model accounting for the nucleon
final state interactions occurring after the weak decay.


\section*{Acknowledgments}
This work has been partially supported by the CONICET,
under contract PIP 6159.


\end{document}